\documentclass[conference,a4paper]{IEEEtran}
\IEEEoverridecommandlockouts 

\usepackage{amsmath,amssymb,enumitem}
\usepackage{graphicx}
\usepackage{tikz}
\usetikzlibrary{graphs} 
\usetikzlibrary{fit} 
\usepackage{microtype}
\usepackage[space]{cite}

\makeatletter
\newcommand\niton{\mathrel{\m@th\mathpalette\canc@l\owns}}
\newcommand\canc@l[2]{{\ooalign{$\hfil#1/\mkern1mu\hfil$\crcr$#1#2$}}}
\makeatother

\newtheorem{theorem}{Theorem}
\newtheorem{corollary}{Corollary}[theorem]
\newtheorem{lemma}{Lemma}
\newtheorem{proposition}{Proposition}
\newtheorem{definition}{Definition}

\newtheorem{remark}{Remark}
\newtheorem{example}{Example}

\graphicspath{{.}{../pictures/}}

\def\gap{1.09ex}
\abovedisplayskip\gap
\belowdisplayskip\gap
\abovedisplayshortskip\gap
\belowdisplayshortskip\gap

\begin{document}

\title{ Secure Index Coding: Existence and Construction\vspace{-0.2ex}}
\author{\IEEEauthorblockN{Lawrence Ong$^1$, Badri N.\ Vellambi$^2$, Phee Lep Yeoh$^3$, J\"{o}rg Kliewer$^2$, and Jinhong Yuan$^4$}
\IEEEauthorblockA{$^1$The University of Newcastle, Australia;\;\; $^2$New Jersey Institute of Technology, USA;\\ $^3$University of Melbourne, Australia;\;\; $^4$University of New South Wales, Australia }
\thanks{This work is supported by ARC grants FT140100219, DE140100420, and DP150100903, and US NSF grants CNS-1526547 and CCF-1439465.}
\vspace{-1ex}
}
\maketitle

\begin{abstract}
We investigate the construction of weakly-secure index codes for a sender to send messages to multiple receivers with side information in the presence of an eavesdropper. We derive a sufficient and necessary condition for the existence of index codes that are secure against an eavesdropper with access to any subset of messages of cardinality $t$, for any fixed $t$. In contrast to the benefits of using random keys in secure network coding, we prove that random keys do not promote security in three classes of index-coding instances.
\end{abstract}


\section{Introduction}
In \textit{classical}\footnote{We use the term classical to indicate the absence of any security constraints.} index-coding problems, a sender sends multiple messages to multiple receivers through a common noiseless broadcast medium, where each receiver has a priori knowledge of a subset of messages~\cite{birkkol2006,baryossefbirk11,blasiakkleinberglubertzky13, yuneely14,arbabjolfaei13}. The subsets that each receiver wants and knows can vary with the receiver.
In this work, we consider \textit{secure} index coding, where in addition to the classical setup, there is an eavesdropper who has access to a subset of messages from a collection of subsets of messages. The sender and the receivers know a priori the collection of message subsets, however, they do not know which subset of messages in this collection is actually accessed by the eavesdropper.   A \textit{weakly-secure} index code must satisfy all receivers' decoding requirements, while ensuring that the eavesdropper is not able to decode any message it has no access to.



\subsection{Contributions of this paper and related work} \label{section:related-work}
The contributions of this work are three-fold:
\subsubsection{Existence of secure index codes}
Secure index coding was first studied by Dau, Skachek, and Chee~\cite{dauskachekchee12}.
The authors derived conditions that any given linear code (of a given message alphabet size) must satisfy to simultaneously meet the receivers' decoding requirements as well as be secure against an eavesdropper with access to a message subset.

In contrast to the code-centric results by Dau et al., we obtain problem-centric results.
We derive a sufficient and necessary condition for the existence of both \textit{linear and non-linear} weakly-secure index codes over \textit{all finite-field alphabets} for any index-coding problem where the eavesdropper can access any message subset of cardinality $t$. We show how to construct such codes if they exist, and investigate their optimality.

\subsubsection{Random keys} 
It has been shown~\cite{caiyeung11} that there exist randomised secure network codes (using random keys) for instances where no deterministic secure network code exists. Owing to an equivalence between classical  versions of network and index coding~\cite{rouayhebsprintsongeorghiades10,effrosrouayheblangberg15}, it is plausible that there exist index-coding instances where randomised encoding can enable security when deterministic encoding cannot. While we do not identify an instance where this is true, we have proven that random keys are \emph{not} useful for weakly-secure index codes in the following three cases: \textit{(i)} the eavesdropper has access to any $t$ messages, \textit{(ii)} the sender's encoding function is linear, or \textit{(iii)} the eavesdropper has access to only one message subset.

\subsubsection{Secure vs classical index coding}
We highlight a significant difference between classical and secure index coding. In classical index coding, messages not required at any receiver are not useful and can be removed from the system. In weakly-secure index coding, these messages may be used as keys.  

\section{Problem Definition and Notation}
Let $m,n\in \mathbb{N}$. For each $i\in [n]\triangleq \{1,\ldots,n\}$, define two subsets $\mathcal{K}_i,\mathcal{W}_i \subseteq [m]$. A classical index-coding instance $(\mathcal K_i, \mathcal W_i)_{i=1}^n$ consists of a single sender and $n$ receivers. The sender has $m$~messages $\boldsymbol{X} = [X_1\, X_2 \,\dotsm X_m]$, where $\{X_i\}_{i=1}^m$  are independent and uniformly distributed over a finite field $\mathcal{F}_q$ with $q$ elements. For a subset of integers $\mathcal{I} = \{i_1, i_2, \dotsc, i_{|\mathcal{I}|}\}$ where $i_1 < i_2 <  \dotsm < i_{|\mathcal{I}|}$, let $\boldsymbol{X}_\mathcal{I} \triangleq [X_{i_1} X_{i_2} \dotsm X_{i_{|\mathcal{I}|}}]$. Each receiver $i\in [n]$ has a priori knowledge of $\boldsymbol{X}_{\mathcal{K}_i}$, and needs to decode $\boldsymbol{X}_{\mathcal W_i}$. The sender encodes $\boldsymbol X$ and gives the codeword to all receivers. The codeword must be chosen so that each receiver $i\in[n]$ is able to decode the messages $\boldsymbol{X}_{\mathcal W_i}$ it wants  using the codeword and the messages $\boldsymbol{X}_{\mathcal K_i}$ it already knows. Without loss of generality, we may assume that $\mathcal{W}_i\setminus \mathcal K_i  \neq \emptyset$  for all $i \in [n]$, since receivers wanting only messages they already know can be expunged from the problem.


Let $\mathfrak{A} \subseteq 2^{[m]}$, where $2^{[m]}$ is the set of all subsets of $[m]$. A secure index-coding instance $((\mathcal{K}_i,\mathcal{W}_i)_{i=1}^n,\mathfrak{A})$ is a classical index-coding instance $(\mathcal{K}_i,\mathcal{W}_i)_{i=1}^n$ in the added presence of an eavesdropper who can access the sender's codeword and precisely one subset of messages $\boldsymbol{X}_{\mathcal A}$, where $\mathcal{A} \in \mathfrak A$. 
The eavesdropper cannot simultaneously access messages corresponding to the indices contained in more than one member of $\mathfrak{A}$. The set $\mathfrak{A}$ contains the possible subsets of indices of compromised messages. While the sender and the receivers are aware of $\mathfrak{A}$, they are oblivious to the exact subset of indices the eavesdropper knows. In addition to meeting the receivers' decoding requirements, a weakly-secure index codeword must ensure that the eavesdropper gains no additional information about each individual message $\boldsymbol{X}_j$, $j\in [m]\setminus \mathcal{A}$, given $\boldsymbol X_\mathcal A$ and the codeword.  Formally, we have the following:

\begin{definition}[Deterministic weakly-secure index code] \label{definition:secure-index-code}
Given a secure index-coding instance $((\mathcal{K}_i,\mathcal{W}_i)_{i=1}^n,\mathfrak{A})$, a deterministic weakly-secure index code $(f_i, \{g_i\}_{i=1}^n)$ of codelength $\ell\in\mathbb{N}$  consists of
\begin{itemize}[leftmargin=*]
\item an encoding function for the sender, $f: \mathcal{F}_q^m \rightarrow  \mathcal{F}_q^\ell$, to encode $\boldsymbol X$ into $\boldsymbol C \triangleq f(\boldsymbol X)$, and
\item a decoding function for each receiver $i\in[n]$, $g_i: \mathcal{F}_q^\ell \times \mathcal{F}_q^{|\mathcal{K}_i|} \rightarrow \mathcal{F}_q^{|\mathcal{W}_i|}$, to decode $\boldsymbol{X}_{\mathcal{W}_i}$ from $\boldsymbol{C}$ and $\boldsymbol{X}_{\mathcal{K}_i}$
\end{itemize}
such that
\begin{itemize}[leftmargin=*]
\item \underline{decodability:} $g_i(f(\boldsymbol{X}), \boldsymbol{X}_{\mathcal{K}_i}) = \boldsymbol{X}_{\mathcal{W}_i}$ for each $i \in [n]$; and
\item \underline{weak security:}  for all $\mathcal{A} \in \mathfrak{A}$, an eavesdropper accessing $\boldsymbol{X}_\mathcal{A}$ has no information about any single message in $\mathcal{A}^\text{c} \triangleq [m]\setminus\mathcal{A}$, i.e., $H(X_i| f(\boldsymbol{X}),\boldsymbol{X}_\mathcal{A}) = H(X_i)$, for all $i \in \mathcal{A}^\text{c}$.$\hfill${\tiny$\blacksquare$}
\end{itemize} 
\end{definition}

\begin{remark} \label{remark:equivalent}
If $\mathfrak{A} = \{[m]\}$, we have a classical index-coding instance without any security constraint.$\hfill${\tiny$\blacksquare$}
\end{remark}

The notion of {weak security} considered here, also known as \textit{$1$-block weakly secure} in the literature~\cite{bhattadnarayanan05, dauskachekchee12},  does not preclude the eavesdropper from gaining information about  $\boldsymbol{X}_{\mathcal{A}^\text{c}}$ despite gaining no knowledge about any single message thereof. 
Other notions of security have also been considered in the literature. For example, Mojahedian, Aref, and Gohari~\cite{mojahedianarefgohari15arxiv} considered \emph{strongly}-secure index coding, where the eavesdropper has \textit{no access} to any message, and must not gain any information about the messages $\boldsymbol X$. Their approach involves the sender encoding messages with keys that are pre-shared with the receivers, but are unknown to the eavesdropper.


It may be possible for the sender to use \emph{random keys} along with the messages $\boldsymbol{X}$ during the encoding process to ensure security against the eavesdropper. We therefore introduce the following notion of random weakly-secure index codes that generalise deterministic weakly-secure index codes. 
\begin{definition}[Random  weakly-secure index code]
Let $Y$ be a random variable taking values in a finite alphabet $\mathcal{Y}$ known only to the the sender, and unknown to the receivers and the eavesdropper. A random weakly-secure index code $(f_i, \{g_i\}_{i=1}^n)$ of codelength $\ell\in\mathbb{N}$  is identical to the deterministic index-code setup with the only exception that the sender encodes $\boldsymbol X$ into $\boldsymbol{C} \triangleq f(\boldsymbol X, Y)$ using the function $f: \mathcal{F}_q^m \times \mathcal{Y} \mapsto \mathcal{F}_q^\ell$. The decoding operations, decodability conditions, and security conditions are identical to those in Definition~\ref{definition:secure-index-code}.
\end{definition}

For the rest of this paper, unless otherwise stated, by secure index codes, we mean weakly-secure index codes.

\begin{definition}[Linear index code]
A random index code is linear if and only if the key $\boldsymbol{Y} = [Y_1\, Y_2 \,\dotsm Y_k]$  for some $k\in\mathbb N$,  where $\{Y_i\}_{i=1}^k$ are independent and uniformly distributed over $\mathcal{F}_q$, and the encoding function 
\begin{align}
\boldsymbol{C} \triangleq f(\boldsymbol{X},\boldsymbol{Y}) = \boldsymbol X \mathbb{G}+ \boldsymbol Y \tilde{\mathbb{G}},
\end{align}
for some matrices $\mathbb{G}$ and $\tilde{\mathbb{G}}$ over $\mathcal F_q$ of sizes $m\times \ell$ and $k\times \ell$, respectively.  Similarly, a deterministic index code is linear if and only if  the encoding function $f(\boldsymbol{X}) = \boldsymbol X \mathbb{G}$. $\hfill${\tiny$\blacksquare$}
\end{definition}


We say that a secure index code exists for a secure index-coding instance $I =((\mathcal{K}_i,\mathcal{W}_i)_{i=1}^n,\mathfrak{A})$ if and only if there exists a (deterministic or random) secure index code $(f,(g_i)_{i=1}^n)$ for some $q$ that meets all the conditions in Definition~\ref{definition:secure-index-code}. If  one such code exists,  we say that the code is \textit{secure against} an eavesdropper having access to any message subset in $\mathfrak{A}$.  As we will see later, a secure index code may or may not exist depending on $\mathfrak{A}$.  The optimal secure index codelength $s(I)$  for a secure index-coding instance $I$, for which secure index codes exist, is defined as infimum of the codelengths of secure index codes over all alphabet sizes.





\section{Fundamental Properties}





We begin with the following counter-intuitive proposition:
\begin{proposition}\label{proposition:security-fewer-messages}
Let $\mathcal{A}' \subsetneq \mathcal{A} \subsetneq [m]$.
An index code secure against an eavesdropper who knows $\boldsymbol{X}_\mathcal{A}$ may not be secure against an eavesdropper who knows $\boldsymbol{X}_{\mathcal{A}'}$, and vice versa.
\end{proposition}

\begin{IEEEproof}
The following example proves this claim. Consider four receivers, where $\mathcal{W}_i = \{i\}$, for all $i \in [4]$,  $\mathcal{K}_1 = \{2\}$, $\mathcal{K}_2 = \{1\}$, $\mathcal{K}_3 = \{2,4\}$, $\mathcal{K}_4 = \{2,3\}$.
Consider two eavesdroppers: the first eavesdropper has access to $\mathfrak{A}_1 = \{\{3,4\}\}$; the second eavesdropper has access to $\mathfrak{A}_2 = \{\{3\}\}$.
The index code $\boldsymbol{C}_1 \triangleq f_1(\boldsymbol{X}) = [X_1 + X_2\;\; X_3+ X_4]$, where $+$ denotes addition of the finite field $\mathcal{F}_q$, is secure against the first eavesdropper (because $H(X_i|\boldsymbol{C}_1,X_3,X_4)=H(X_i)$ for each $i \in\{1,2\}$) but not the second eavesdropper (because it can decode $X_4$). The index code $\boldsymbol{C}_2 \triangleq f_2(\boldsymbol{X}) = [X_1 + X_2\;\; X_2 + X_3 + X_4]$ is secure against the second eavesdropper, but not the first. 
\end{IEEEproof}


Proposition~\ref{proposition:security-fewer-messages} is in contrast to secure network coding~\cite{caiyeung11}, where a network code that is strongly secure against an eavesdropper who can access a subset of links, say $\mathcal{L}$, is also secure against an eavesdropper who can access any $\mathcal{L}' \subsetneq \mathcal{L}$.

\begin{proposition} \label{proposition:more-knowledge}
No secure index code exists for any secure index-coding instance $((\mathcal{K}_i,\mathcal{W}_i)_{i=1}^n,\mathfrak{A})$ where there exists $\mathcal{A} \in \mathfrak{A}$ and $i\in[n]$ such that 
$\mathcal{K}_i \subseteq \mathcal{A}$ and $\mathcal{W}_i \cap \mathcal{A}^\text{c} \neq \emptyset$.
\end{proposition}

\begin{IEEEproof}
Pick $j \in \mathcal{W}_i \cap \mathcal{A}^\text{c}$. Let $\boldsymbol{C}=f(\boldsymbol{X},Y)$ be the codeword of a random index code that uses a key $Y$. Since  $H(X_j|\boldsymbol{C},\boldsymbol{X}_{\mathcal{A}}) \leq H(X_j|\boldsymbol{C},\boldsymbol{X}_{\mathcal{K}_i})=0$, the eavesdropper is able to decode $\boldsymbol X_j$. Thus, the code cannot be secure.
\end{IEEEproof}


\section{Existence of Secure Index Codes}

Here, we present a necessary and sufficient condition for the existence of secure index codes, and their construction. Furthermore, we derive optimal secure index codes for certain classes of instances. We begin with a specific type  of eavesdroppers.

%
\begin{definition}
For a given $(\mathcal{K}_i, \mathcal{W}_i)_{i=1}^n$,
we say that an index code is \textit{secure against an eavesdropper with $t$-level access}, for some $t \in \{0,1,\dotsc, m-1\}$, if and only if it is a secure index code for $((\mathcal{K}_i, \mathcal{W}_i)_{i=1}^n, \{\mathcal{A} \subsetneq [m]: |\mathcal{A}|=t\})$.$\hfill${\tiny$\blacksquare$}
\end{definition}

\begin{lemma}\label{lemma:weaker}
Any (deterministic or random) index code secure against an eavesdropper with $t$-level access is also secure against an eavesdropper with $t'$-level access, for any $t'<t$.
\end{lemma}

\begin{IEEEproof}
Consider an eavesdropper with $t$-level access who has access to any member of $\mathfrak{A} = \{\mathcal{A}\subsetneq [m]: |\mathcal{A}|=t\}$. An index code secure against this eavesdropper must satisfy 
\begin{equation}
H(X_i | \boldsymbol{C},\boldsymbol{X}_\mathcal{A}) = H(X_i), \quad \text{for all } \mathcal A \in\mathfrak A, i \in \mathcal{A}^\text{c}. \label{eq:secure-t}
\end{equation}
%
Consider an eavesdropper with an access level $t' < t $ who has access to any member of $\mathfrak{A}' = \{\mathcal{B}\subsetneq [m]: |\mathcal{B}|=t'\}$.
Pick any $\mathcal{A}'  \in \mathfrak{A}'$ and any  $i \in [m] \setminus \mathcal{A}'$. As $t'< t \leq m-1$, we can always find a subset  $\mathcal{A} \in \mathfrak{A}$ such that $\mathcal{A}' \subsetneq \mathcal{A}$ and $i \notin\mathcal{A}$. So,
\begin{equation*}
H(X_i) \stackrel{(a)}{=} H(X_i|\boldsymbol{C},\boldsymbol{X}_\mathcal{A}) \stackrel{(b)}{\leq} H(X_i|\boldsymbol{C},\boldsymbol{X}_{\mathcal{A}'})  \stackrel{(c)}{\leq} H(X_i),
\end{equation*}
where $(a)$ follows from \eqref{eq:secure-t}, and $(b)$ and  $(c)$ follow since conditioning cannot increase entropy. Since the choices of $\mathcal{A}'$ and $i$ are arbitrary, we must have $H(X_i | \boldsymbol{C},\boldsymbol{X}_{\mathcal{A}'}) = H(X_i)$, for all  $\mathcal{A}' \in \mathfrak{A}'$ and all $i \in [m] \setminus \mathcal{A}'$. Thus, the index code is also secure against an eavesdropper with $t'$-level access.
\end{IEEEproof}

\begin{remark}
Lemma~\ref{lemma:weaker} generalises the result by Dau et al.~\cite[Theorem~4.9]{dauskachekchee12} that pertains specifically to deterministic linear index codes to any (random or deterministic, linear or non-linear) index code.$\hfill${\tiny$\blacksquare$}
\end{remark}

\begin{remark}
Although Proposition~\ref{proposition:security-fewer-messages} states that an index code  secure against an eavesdropper with access to $\mathfrak{A}=\{\mathcal{A}\}$ may not be secure against an eavesdropper with access to $\mathfrak{A}=\{\mathcal{A}'\}$ where $\mathcal{A}' \subsetneq \mathcal{A}$, any index code secure against an eavesdropper with $t$-level access, i.e., $\mathfrak{A} = \{\mathcal{A} \subsetneq [m] :|\mathcal A| = t\}$, \textit{is also}  secure against an eavesdropper with any access level $t' < t$.$\hfill${\tiny$\blacksquare$}
\end{remark}

\subsection{Existence of secure index codes and their construction}

We now present a necessary and sufficient condition for the existence of secure index codes.

\begin{theorem} \label{thm:necessary-sufficient}
Consider a secure index-coding instance $((\mathcal{K}_i, \mathcal{W}_i)_{i=1}^n,\mathfrak{A})$ with $\mathfrak{A} = \{\mathcal{A} \subsetneq [m] :|\mathcal A| = t\}$ for some $t < m$,
i.e., the eavesdropper has $t$-level access.
Secure index codes exist if and only if
\begin{equation}
t <  K_\text{min} \triangleq \min_{i \in [n]} |\mathcal{K}_i|. \label{eq:necessary-sufficient-existence}
\end{equation}
Deterministic linear secure index codes exist if \eqref{eq:necessary-sufficient-existence} is satisfied.
\end{theorem}

\begin{IEEEproof}
We first prove the converse. Suppose that $t \geq K_\text{min}$. By definition, there exists a receiver, say $i$, with $|\mathcal{K}_i| = K_\text{min}$, and $\mathcal{W}_i\setminus \mathcal{K}_i \neq \emptyset$. Pick some $j \in \mathcal{W}_i\setminus \mathcal{K}_i$, i.e., $\mathcal{K}_i \subseteq [m] \setminus \{j\}$. Since $K_\text{min} \leq t \leq m-1$, we can always find some $\mathcal{A} \in  \mathfrak{A}$ such that $\mathcal{K}_i \subseteq \mathcal{A} \subseteq [m] \setminus \{j\}$. From Proposition~\ref{proposition:more-knowledge}, we conclude that no secure index code exists.


Next, we prove the forward part. 
Consider a deterministic  linear index code of length $\ell = m - K_\text{min}$, formed by
\begin{equation}
\boldsymbol{C} = \boldsymbol{X} \mathbb{G} =
\sum_{i \in [m]} X_i \boldsymbol{g}_i, \label{eq:mds}
\end{equation}
where $\mathbb{G}$ is an $m\times \ell$ matrix over $\mathcal{F}_q$, and $\boldsymbol{g}_i \in \mathcal{F}_q^\ell$ is the $i$-th row of $\mathbb{G}$. Let $\mathbb{G}$ be the \textit{transpose} of the generating matrix of a maximum-distance-separable (MDS) code, which always exists for a sufficiently large $q$. For any such code, it follows that any $\ell$ rows of $\mathbb{G}$ are linearly independent over $\mathcal{F}_q$.

(Decoding) Receiver $i \in [n]$ forms
\begin{equation}
\boldsymbol{C} - \sum_{k \in \mathcal{K}_i} X_k \boldsymbol{g}_k = \sum_{j \in [m] \setminus \mathcal{K}_i} X_j \boldsymbol{g}_j. \label{eq:decoding}
\end{equation}
Since $|[m] \setminus \mathcal{K}_i| = m - |\mathcal{K}_i| \leq m - K_\text{min} = \ell$, it follows that $\{\boldsymbol{g}_j:j \in [m] \setminus \mathcal{K}_i\}$ are  linearly independent. So,  by using $\boldsymbol{C}$ and $\{X_k: k \in \mathcal{K}_i\}$,  receiver $i$ can decode  $\boldsymbol{X}_{[m] \setminus \mathcal{K}_i}$, and consequently the message(s) $\boldsymbol{X}_{\mathcal W_i}$ it wants, by solving \eqref{eq:decoding}. 

(Security) Denote the Hamming distance between two vectors $\boldsymbol{a} = [a_1 a_2 \dotsm a_m] \in \mathcal{F}_q^m$ and $\boldsymbol{b} = [b_1  b_2  \dotsm b_m] \in \mathcal{F}_q^m$ by $d(\boldsymbol{a}, \boldsymbol{b}) \triangleq |\{ i \in [m]: a_i \neq b_i\}|$, the minimum distance of a vector space $\mathcal{S}$ by
$d(\mathcal{S}) = \min_{\{\boldsymbol{a}, \boldsymbol{b} \in \mathcal{S}: \boldsymbol{a} \neq \boldsymbol{b}\}} d(\boldsymbol{a}, \boldsymbol{b})$,
and the vector space spanned by the rows and columns of a matrix $\mathbb{M}$ by $\mathsf{rowsp}(\mathbb{M})$ and $\mathsf{colsp}(\mathbb{G})$, respectively. 
Dau et al.~\cite{dauskachekchee12} showed that the linear index code of the form~\eqref{eq:mds} is secure against an eavesdropper with an access level $d(\mathsf{colsp}(\mathbb{G}))-2$. Note that $\mathbb{G}^\text{T}$ is the generator matrix of an MDS code $(m, \ell, d)$ whose codewords are vectors in  $\mathsf{rowsp}(\mathbb{G}^\text{T})$. The minimum distance of this MDS code equals $d = d(\mathsf{rowsp}(\mathbb{G}^\text{T})) = d(\mathsf{colsp}(\mathbb{G})) = m - \ell +1$. Invoking Lemma~\ref{lemma:weaker}, we see that the index code~\eqref{eq:mds} is secure against an eavesdropper with an access level up to and including $(m - \ell + 1 ) - 2 = K_\text{min} -1$.
\end{IEEEproof}
Some remarks are now in order.
\begin{remark}
MDS codes are also used in the partial-clique-cover coding scheme~\cite{birkkol2006} and its time-shared version~\cite{yuneely14}, and the local-chromatic-number coding scheme~\cite{shanmugamdimakislangberg13} for \textit{unicast} index coding, where $\mathcal{W}_i=\{i\}$ for all receivers $i \in [n]$.$\hfill${\tiny$\blacksquare$}
\end{remark}

\begin{remark}
Receiver cooperation can increase the security level. Allowing two receivers, say $i$ and $j$, to cooperate and share their messages is equivalent to solving a new secure index-coding instance where everything remains the same except that receivers $i$ and $j$ both know $\boldsymbol{X}_{\mathcal{K}_i \cup \mathcal{K}_j}$. Thus, cooperation can potentially increase $K_{\min}$ (see \eqref{eq:necessary-sufficient-existence}), which then translates to security against eavesdroppers with higher access levels.$\hfill${\tiny$\blacksquare$}
\end{remark}

\begin{remark} \label{remark:b-block}
Theorem~1 also holds if we consider $b$-block security (see Dau et al.~\cite{dauskachekchee12}) for $b\geq 1$ in Definition~\ref{definition:secure-index-code}. In the setting of $b$-block security, an eavesdropper who knows $\boldsymbol{X}_\mathcal{A}$, $\mathcal A \in \mathfrak A$, gains no information about any  $b$ messages it  does not know, i.e., $H(\boldsymbol{X}_\mathcal{B}|\boldsymbol{C},\boldsymbol{X}_\mathcal{A}) = H(\boldsymbol{X}_\mathcal{B})$, for all $\mathcal{B} \subseteq \mathcal{A}^\text{c}$ with $|\mathcal{B}| = b$. In this case, the necessary and sufficient condition for the existence of secure index codes in \eqref{eq:necessary-sufficient-existence}  is replaced by $t \leq K_\text{min} - b$.$\hfill${\tiny$\blacksquare$}
\end{remark}


\begin{corollary}\label{cor:sufficient-condition}
If $A_\text{max} \triangleq \max_{\mathcal{A} \in \mathfrak{A}} |\mathcal{A}| <  K_\text{min} $,  then deterministic linear secure index codes exist.
\end{corollary}

\begin{IEEEproof}
Proof follows from Theorem~\ref{thm:necessary-sufficient} and Lemma~\ref{lemma:weaker}.
\end{IEEEproof}

Intuitively, Corollary~\ref{cor:sufficient-condition} says that we can always find secure index codes if the eavesdropper can access fewer messages than each receiver can. However, unlike Theorem~\ref{thm:necessary-sufficient}, we do not have a converse for Corollary~\ref{cor:sufficient-condition}. This is because even if an eavesdropper can access (numerically) more messages than some receivers can, we may still be able to construct  secure  index codes, depending on the sets of messages to which the eavesdropper has access.  For example, see the secure index codes for the two instances with $A_\text{max} \geq K_\text{min}$ in the proof of Proposition~\ref{proposition:security-fewer-messages}.


\subsection{Optimality of secure index codes}

From the construction of secure index codes in Theorem~\ref{thm:necessary-sufficient}, we have the following:
\begin{corollary} \label{cor:upper bound}
If $A_\text{max} < K_\text{min}$, the optimal secure index codelength is
 upper-bounded as $s(I) \leq m - K_\text{min}$.
The upper bound is achievable by deterministic linear index codes.
\end{corollary}

\begin{IEEEproof}
See the proof of Theorem~\ref{thm:necessary-sufficient}.
\end{IEEEproof}


We say that a receiver $i$  has \textit{complementary message requests} if it wants all messages it does not know, i.e., $\mathcal{K}_i \cup \mathcal{W}_i = [m]$.

\begin{proposition} \label{cor:complementary}
If $A_\text{max} < K_\text{min}$, and if any receiver knowing  exactly $K_\text{min}$ messages has complementary message requests, then the optimal codelength $s(I) = m - K_\text{min}$, and 
is achievable by deterministic secure linear index codes. 
\end{proposition}

\begin{IEEEproof}
Without loss of generality, let $|\mathcal K_1|= K_{\min}$ and $\mathcal K_1 \cup \mathcal W_1 = [m]$. For any (deterministic or random) index code $\boldsymbol{C}$, we have $H(\boldsymbol{X}_{\mathcal{W}_1} | \boldsymbol{C},\boldsymbol{X}_{\mathcal{K}_1})=0$.  Thus,
\begin{subequations}
\begin{align*}
m \log_2 q &= H(\boldsymbol{X}) \stackrel{}{\leq} H(\boldsymbol{C},\boldsymbol{X}_{\mathcal{K}_1},\boldsymbol{X}_{\mathcal{W}_1}) \stackrel{}{=}  H(\boldsymbol{C},\boldsymbol{X}_{\mathcal{K}_1})\\
&\stackrel{}{\leq}\log_2 q^\ell + \log_2 q^{K_{\min}},
\end{align*}
\end{subequations}
where we have made use of the facts that $ \boldsymbol{X}= \boldsymbol{X}_{\mathcal{K}_1\cup \mathcal{W}_1}$,  $H(\boldsymbol{X}_{\mathcal{W}_1} | \boldsymbol{C},\boldsymbol{X}_{\mathcal{K}_1})=0$, $H(\boldsymbol{C}) \leq \log_2 |\boldsymbol C| = \log_2 q^\ell $, and $H(\boldsymbol{X}_{\mathcal{K}_1}) = \log_2 q^{K_{\min}}$. Therefore, $\ell \geq m - K_\text{min}$. Since $\boldsymbol{C}$ was arbitrary, it follows that $s(I) = \inf \ell \geq m - K_\text{min}$. The proof is then complete by invoking Corollary~\ref{cor:upper bound}.
\end{IEEEproof}

\section{Secure vs Classical Index Coding}

We can represent a secure index-coding instance $((\mathcal{G}_i,\mathcal{K}_i)_{i=1}^n,\mathfrak{A})$ by a directed bipartite graph $\mathcal{D}=(\mathcal{U}, \mathcal{M}, \mathcal{E})$, similar to that by Neely, Tehrani, and Zhang~\cite{neelytehranizhang12}. Here, $\mathcal{U}$ and $\mathcal{M}$ are independent vertex sets, where each arc (i.e., directed edge) in $\mathcal{E}$ connects a vertex in $\mathcal{U}$ to a vertex in $\mathcal{M}$. We further partition $\mathcal{U}$ into two disjoint sets: $\mathcal{R} = \{r_1, r_2, \dotsc, r_n\}$ representing the $n$ receivers, and $\mathcal{V} = \{v_1, v_2, \dotsc, v_{|\mathfrak{A}|}\}$ representing the possible sets of messages to which the eavesdropper can access. The set $\mathcal{M} = [m]$ represents the message indices. The arc set $\mathcal{E}$ is defined as follows:
\begin{itemize}
\item There is an arc from $r_i \in \mathcal{R}$ to $j \in \mathcal{M}$ if and only if receiver~$i$ knows the message~$X_j$, i.e., $j \in \mathcal{K}_i$.
\item There is an arc from $j \in \mathcal{M}$ to $r_i \in \mathcal{R}$ if and only if receiver~$i$ wants the message~$X_j$, i.e., $j \in \mathcal{W}_i$.
\item For each $\mathcal{A} \in \mathfrak{A}$, we have a unique $v_i \in \mathcal{V}$ such that $\mathcal{N}^+_{\mathcal{D}}(v_i) = \mathcal{A}$, where $\mathcal{N}^+_{\mathcal{D}}(v_i) \triangleq \{ j \in \mathcal{M}: (v_i \rightarrow j) \in \mathcal{E}\}$ is the out-neighbourhood of $v_i$.
\end{itemize}


For a given secure index-coding instance $\mathcal{D}$, if we ignore the security constraint, the subgraph $\mathcal{D}[\mathcal{R} \cup \mathcal{M}]$ induced by $(\mathcal{R},\mathcal{M})$ is in fact the bipartite graph used by Neely et al.~\cite{neelytehranizhang12} to represent the classical index-coding instance.

\begin{proposition} \label{proposition:acyclic}
Consider a secure index-coding instance $((\mathcal{G}_i,\mathcal{K}_i)_{i=1}^n,\mathfrak{A})$, where $\mathfrak{A} \neq \{[m]\}$. Let $\mathcal{D}=((\mathcal{R},\mathcal{V}),\mathcal{M},\mathcal{E})$ be its directed bipartite graph representation. If
\begin{itemize}
\item[(C1)]  $\mathcal{D}[\mathcal{R} \cup \mathcal{M}]$ is acyclic, or equivalently, $\mathcal{D}$ is acyclic, and
\item[(C2)] every message is wanted by some receiver, i.e., for each $i \in [m]$, we have $i \in \mathcal{W}_j$ for some $j \in [n]$,
\end{itemize}
 then no secure index code exists.
\end{proposition}

\begin{IEEEproof}
For the classical index-coding instance $\mathcal{D}[\mathcal{R} \cup \mathcal{M}]$, Neely et al.~\cite[Appendix~A]{neelytehranizhang12} have shown that if condition C1 is true (condition C2 is always assumed to be true for non-secure index coding), one can obtain all messages from any index code, even without using side information. Since any secure index code for $\mathcal{D}$, denoted by $\boldsymbol{C}$, is an index code for $\mathcal{D}[\mathcal{R} \cup \mathcal{M}]$, we have $H(\boldsymbol{X}_{[m]}|\boldsymbol{C}) = 0$. Therefore, for any $\mathcal{A} \subsetneq [m]$ and any $i \in [m] \setminus \mathcal{A}$, we have
$H(X_i|\boldsymbol{C},\boldsymbol{X}_{\mathcal{A}}) \leq   H(\boldsymbol{X}_{[m]}|\boldsymbol{C})= 0 < H(X_i)$.  
Since $\mathfrak{A} \neq \{[m]\}$, there exists some $\mathcal{A} \subsetneq [m]$. It follows  that no index code can be secure.
\end{IEEEproof}


Condition C2 in Proposition~\ref{proposition:acyclic} that every message is wanted by some receiver is implicit in classical index coding as removing messages not wanted by any receiver will change neither the index code nor the optimal index codelength. However, removing unwanted messages may affect secure index coding, because these messages can be used as keys to protect the index code against the eavesdropper. The following example illustrates this idea.
\begin{example} \label{example:unwated-messages}
Consider the following secure index-coding instance depicted by its directed bipartite graph representation.
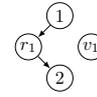
\begin{center}
\resizebox{8ex}{!}{%
\begin{tikzpicture}
\graph {[math nodes, nodes={draw,circle,inner sep = 0, minimum size=5mm},edge={>=latex} ,clockwise = 4, chain polar shift=(0:1.5cm), radius=0.6cm]
1, v_1, 2, r_1,
1 -> r_1,
r_1 -> 2,
};
\end{tikzpicture}
}%
\end{center}
The message $X_2$ is not wanted by any receiver. If we remove it from the setup, by invoking Proposition~\ref{proposition:acyclic}, we conclude that there is no secure index code. However, keeping $X_2$ in the system, by invoking Corollary~\ref{cor:sufficient-condition}, we conclude that secure index codes exist. Indeed, the index code $C = X_1 + X_2$ is secure. Here, $X_2$ acts as a key between the sender and receiver~1 to protect message $X_1$ against the eavesdropper.
\end{example}

\section{Random Keys for Secure Index Coding}

We saw in Example~\ref{example:unwated-messages} that using unwanted messages as keys may be  essential in ensuring security. One wonders if generating random keys unknown to the receivers and the eavesdropper can also help in secure index coding. 
While the answer to this question is not known in general, we show that in the following three scenarios, random keys are not useful in the sense that random secure index codes exist if and only if deterministic secure index codes also exist.

\subsection{Eavesdroppers with $t$-level access}

From Theorem~1, it follows that using random keys does not provide greater security against an eavesdropper with $t$-level access, i.e., when $\mathfrak{A} = \{\mathcal{A} \subsetneq [m] :|\mathcal A| = t\}$, for any $t < m$. 


\subsection{Linear index codes}
We now restrict the secure index codes to be linear, while $\mathfrak{A}$ is arbitrary.
\begin{theorem}
Given any secure index-coding instance $I$.
Random secure \textit{linear} index codes of codelength $\ell$ exist for $I$ if and only if deterministic secure \textit{linear} index codes of codelength $\ell$ also exist for $I$.
\end{theorem}
\begin{IEEEproof}
We only need to prove the only if direction of the claim. Any random  linear index code can be expressed as $\boldsymbol{C} = \boldsymbol X \mathbb{G}+ \boldsymbol Y \tilde{\mathbb{G}}$.
Since each receiver recovers its intended messages, for each receiver $i \in [n]$ and each $j\in \mathcal W_i$, there exist an $\ell\times 1$ vector $\mathbf D_{i,j}$ and a $|\mathcal{K}_i| \times 1$ vector $\mathbf E_{i,j}$ such that
\begin{align}
X_j = \boldsymbol C \mathbf D_{i,j} + \boldsymbol X_{\mathcal K_i} \mathbf E_{i,j}. \label{eqn:LinDecEqns}
\end{align}
Let $\mathbb V$ be defined as the nullspace of $\tilde{\mathbb{G}}$, i.e.,
\begin{align}
\mathbb{V} =\mathsf{Null}(\tilde{\mathbb{G}}) \triangleq \{ \mathbf{A}\in \mathcal F_q^\ell : \tilde{\mathbb{G}}\mathbf{A} = \boldsymbol{0}\}.
\end{align}
Note that $\mathbb{V}$ is a vector space. From \eqref{eqn:LinDecEqns}, it follows that $\mathbf D_{i,j} \in \mathbb V$ for any $i\in[n]$ and $j\in \mathcal W_i$, since
\begin{align}
X_j - \boldsymbol X_{\mathcal K_i} \mathbf E_{i,j}  = \boldsymbol C \mathbf D_{i,j} =\boldsymbol X \mathbb{G} \mathbf D_{i,j} + \boldsymbol Y \tilde{\mathbb G} \mathbf D_{i,j},  \end{align}
which can hold only if $ \tilde{\mathbb G} \mathbf D_{i,j} = \mathbf 0$ for any $i\in[n]$ and $j\in \mathcal W_i$.

Now, let $\mathbf{A}_1,\ldots,\mathbf{A}_{\hat \ell}$ be a basis for $\mathbb V$. Note that $\hat l \leq l $, since  $\mathbb V\subseteq \mathcal F_q^\ell$. If the sender broadcasts $\hat{\boldsymbol{C}} \triangleq [ \hat C_1\,\,\hat C_2\,\, \cdots\,\, \hat C_{\hat \ell} ]$, where $\hat C_i = \boldsymbol C\mathbf A_i = \boldsymbol X \mathbb{G} \mathbf A_i$, $i \in [\hat{\ell}]$, then each receiver will still be able to recover its intended messages, since for any $i\in[m]$ and $j\in \mathcal W_i$, $ X_j - \boldsymbol X_{\mathcal K_i} \mathbf E_{i,j}=\hat{\boldsymbol{C}} \mathbf D_{i,j}$ is a linear combination of $\hat C_1,\ldots, \hat C_{\hat \ell}$. Furthermore, for any $\mathcal{A}\in \mathfrak A$ and any $j\in \mathcal{A}^\text{c}$,
\begin{align}
\hspace{-2mm}H(X_j) \geq H(X_j | \boldsymbol{X}_\mathcal A, \hat{\boldsymbol{C}}) \geq H(X_j | \boldsymbol{X}_\mathcal A, {\boldsymbol{C}}) = H(X_j),
\end{align}
where the second inequality follows since $\hat{\boldsymbol{C}}$ is a function of $\boldsymbol{C}$. Hence, the new code $\hat{\boldsymbol{C}}$ is also secure. The proof is then complete by noting that $\hat{\boldsymbol{C}}$ is a deterministic index code.
\end{IEEEproof}

\begin{remark}
Random keys have also been shown  not to be useful for linear secure index codes in the strong-security setting considered by Mojahedian et al.~\cite{mojahedianarefgohari15arxiv}.
\end{remark}

\subsection{Eavesdroppers having access to only one message subset}
Lastly, we consider the class of secure index-coding instances where the eavesdropper can access only one message subset.

\begin{proposition}
Given any index-coding instance $I$
with $|\mathfrak{A}|=1$,  random  secure index codes exist for $I$ if and only if deterministic secure index codes also exist for $I$.
\end{proposition}

\begin{IEEEproof}
Note that we only need to consider index-coding instances where for each receiver $i \in [n]$, either
\begin{itemize}
\item (Type 1) $\mathcal{K}_i \cup \mathcal{W}_i \subseteq \mathcal{A}$;
\item (Type 2) $\mathcal{K}_i \setminus \mathcal{A} \neq \emptyset$ and $\mathcal{W}_i \setminus \mathcal{A} \neq \emptyset$; or
\item (Type 3) $\mathcal{K}_i \setminus \mathcal{A} \neq \emptyset$ and $\mathcal{W}_i \subseteq \mathcal{A}$.
\end{itemize}
Otherwise, according to Proposition~\ref{proposition:more-knowledge}, no (deterministic or random) secure index code exists. Let receivers $1,\dotsc, n'$ be of Type~2, for some $ 0\leq n' \leq n$, and the rest, of Type~1 or 3.

Consider a related index-coding  instance $I'=((\mathcal{K}'_i,\mathcal{W}'_i)_{i=1}^{n'}, \mathfrak{A}')$  with only $|\mathcal A|^c$ messages $\boldsymbol{X}_{\mathcal{A}^\text{c}}$, $n'$ receivers that are of Type~2 in $I$, where $\mathfrak{A}' = \{\emptyset\}$,  $\mathcal{K}'_i = (\mathcal{K}_i \cap \mathcal{A}^\text{c}) \neq \emptyset$, and $\mathcal{W}'_i = (\mathcal{W}_i \cap \mathcal{A}^\text{c}) \neq \emptyset$, for all $i \in [n']$.  By the definition of Type-2 receiver, $|\mathcal{K}'_i| \geq 1$ for all $i \in [n']$. For $I'$, as $A_\text{max} = 0$ and $K_\text{min} \geq 1$, by invoking Corollary~\ref{cor:sufficient-condition}, we see that there exists a deterministic secure index code, say, $\boldsymbol{C}' = f'(\boldsymbol{X}_{\mathcal{A}^\text{c}})$. This means that there exists a function $g'_i(\boldsymbol{C}',\boldsymbol{X}_{\mathcal{K}'_i}) = \boldsymbol{X}_{\mathcal{W}'_i}$ for each $i \in [n']$, and $H(X_i|\boldsymbol{C}') = H(X_i)$ for each $i \in \mathcal{A}^\text{c}$.

We now show that $\boldsymbol{C} = [\boldsymbol{X}_{\mathcal{A}} \,\, \boldsymbol{C}']$ is a secure index code for $I$. For any receiver of Type~1 or 3, its decoding requirement is fulfilled from observing $\boldsymbol{X}_{\mathcal{A}}$, because $\mathcal{W}_i \subseteq \mathcal{A}$. Any receiver~$i$ of Type~2 gets $\boldsymbol{X}_{\mathcal{W}_i \cap \mathcal{A}}$ from $\boldsymbol{X}_\mathcal{A}$, and $\boldsymbol{X}_{\mathcal{W}_i \cap \mathcal{A}^\text{c}} = \boldsymbol{X}_{\mathcal{W}'_i}$ from $g'_i(\boldsymbol{C}',\boldsymbol{X}_{\mathcal{K}'_i})$, since it knows $\boldsymbol{X}_{\mathcal{K}_i} \supseteq \boldsymbol{X}_{\mathcal{K}'_i}$.

Finally, $H(X_i | \boldsymbol{C}, \boldsymbol{X}_\mathcal{A}) = H(X_i | \boldsymbol{C}', \boldsymbol{X}_\mathcal{A}) \stackrel{(a)}{=} H(X_i | \boldsymbol{C}') = H(X_i)$, for  any  $i \in \mathcal{A}^\text{c}$, where $(a)$ follows from the independence of $(X_i,\boldsymbol{C}')$ and $\boldsymbol{X}_\mathcal{A}$. Hence, $\boldsymbol{C}$ is a deterministic secure index code for $I$. So, for any $I$ with $|\mathfrak{A}| = 1$, either no (deterministic or  random) secure index codes exist, or we can always find a deterministic secure index code.
\end{IEEEproof}

\subsection{Secure index vs network coding}


We  now discuss some  issues in extending the equivalence\footnote{The instances are equivalent in the sense that a code for one instance can be translated to a code  for the other, and vice versa.}~\cite{rouayhebsprintsongeorghiades10,effrosrouayheblangberg15} between  classical  index and network  coding to the secure setting. Consider the following network-coding instance $N$ with a source $s$ having two links  $e_1$ and $e_2$  to a receiver $d$.
\vspace*{-1ex}
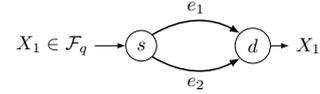
\begin{center}
\resizebox{27ex}{!}{%
\begin{tikzpicture}
\def\D{20mm}

\node[draw, circle] (S) at (0,0) {$s$};
\node[draw, circle] (D) at (\D,0) {$d$};
\draw[thick, ->, >=latex, bend left] (S.north east) to node[midway,above] {$e_1$} (D.north west) ;
\draw[thick, ->, >=latex, bend right] (S.south east) to node[midway,below] {$e_2$} (D.south west) ;
\node (X) at (-\D*0.8,0) {$X_1 \in \mathcal{F}_q$};
\node (X2) at (\D*1.5,0) {$X_1$};
\draw[->, >=latex] (X) to (S);
\draw[->, >=latex] (D) to (X2);
\end{tikzpicture}
}%
\end{center}
\vspace*{-2ex}
The codewords conveyed on links $e_1$ and $e_2$ in any network code can be  written as $Y_1 = f_1(X_1)$ and $Y_2=f_2(X_1)$, respectively, and the decoding operation $ X_1 = g(Y_1,Y_2)$. 

An equivalent~\cite{effrosrouayheblangberg15}  index-coding  instance $I$ has three  independent messages $\hat{X}_1, \hat{Y}_1, \hat{Y}_2$, and four receivers, as follows: 
\vspace*{-3ex}
\begin{center}
\begin{scriptsize}
\begin{tabular}{c| c c c c}
Receiver & 1 & 2 & 3 & 4 \\
\hline
Has & $\hat{X}_1$ & $\hat{X}_1$ & $(\hat{Y}_1,\hat{Y}_2)$ & $\hat{X}_1$\\
Wants & $\hat{Y}_1$ & $\hat{Y}_2$ & $\hat{X}_1$ & $(\hat{Y}_1,\hat{Y}_2)$
\end{tabular}
\end{scriptsize}
\end{center}
\vspace*{-0.5ex}
Since the codes for instances $I$ and $N$ can be translated to one another~\cite{effrosrouayheblangberg15}, we can translate the above code for $N$ to an  index code   
$\hat{\boldsymbol{C}} = [ \hat{Y}_1 + f_1(\hat{X}_1) \;\; \hat{Y}_2 + f_2(\hat{X}_1)]$ for $I$. 

Next, consider a secure version of $N$, with an eavesdropper who has access to any one link ($e_1$ or $e_2$)~\cite{caiyeung11}. A secure network code must strongly secure $X_1$ against the eavesdropper. To this end, we need \textit{random} network codes, e.g., $f_1(X_1) = K$, where $K$ is a random key uniformly distributed on $\mathcal{F}_q$ and independent of $X_1$, and $f_2(X_1) = X_1 + K$.

Unfortunately, the code translation breaks down here in the presence of security constraints. In $I$, for receiver~1 to decode $\hat{Y}_1$ from $\hat{\boldsymbol{C}}$ and the message $\hat{X}_1$ it knows, it additionally needs to know the random key $K=f_1(\hat{X}_1)$ generated by the sender. 

One difficulty in establishing an equivalence between secure network coding and secure index coding is that random keys used by the sender for encoding need not be available to the receivers for decoding. Furthermore, it is also not straightforward to translate (strong or weak) security constraints for the eavesdropper in $N$ to equivalent and meaningful (strong or weak) security constraints in $I$, and vice versa.

\end{document}